\documentclass[%
superscriptaddress,
twocolumn,
 amsmath,amssymb,
 aps,
prb,
floatfix,
]{revtex4-2}

\usepackage{graphicx}
\usepackage{dcolumn}
\usepackage{bm}
\usepackage{xcolor}
\usepackage{multirow}
\usepackage{array}
\usepackage{ulem}
\definecolor{revised}{RGB}{0, 0, 0}

\begin{document}

\preprint{APS/123-QED}

\title{
Laser-driven first-order spin reorientation and Verwey
phase transitions\\ in the magnetite Fe$_3$O$_4$
beyond the range of thermodynamic equilibrium
}
\author{A. V. Kuzikova}
\email{anna.kuzikova@mail.ioffe.ru}
\author{L. A. Shelukhin}%
\affiliation{%
 Ioffe Institute, 194021 St. Petersburg, Russia}%

\author{F. M. Maksimov}
\author{A. I. Chernov}
\affiliation{%
 Russian Quantum Center, 143025, Skolkovo,  Russia}%
 \affiliation{%
 Center for Photonics and 2D Materials, Moscow Institute of Physics and Technology, 14170, Dolgoprudny, Russia}%
\author{R. V. Pisarev}
\author{A. M. Kalashnikova}
\affiliation{%
 Ioffe Institute, 194021 St. Petersburg, Russia}

\date{\today}

\begin{abstract}
Ultrafast photo-induced phase transitions occurring under the impact of femtosecond laser pulses provide versatile opportunities for switching solids between distinctly-different crystalline, electronic, and magnetic states and thus modify their functional properties in a significant way.
In this paper, we report on the laser-induced spin reorientation and Verwey phase transitions in a single crystalline ferrimagnetic magnetite Fe$_3$O$_4$. 
Using femtosecond optical and magneto-optical pump-probe techniques, we define the range of the initial sample temperatures and laser fluences when partial or complete photo-induced phase transitions occur from a monoclinic insulating to a cubic metallic state with concomitant switching of magnetic anisotropy from the uniaxial to the cubic one.  
We thus reveal a connection between these phase transitions when driven by femtosecond laser pulses.
Using transient linear and quadratic magneto-optical effects, we examine magnetization dynamics launched the switching of the magnetic anisotropy axis. We unveil the presence of the domains undergoing the laser-induced phase transitions even below the established threshold fluence for the transitions, as well as when the material is initially in the cubic phase. This is the manifestation of the first-order of these both laser-induced phase transitions beyond the range of thermodynamic equilibrium.
\end{abstract}

\maketitle
\section{Introduction} \label{sec:intro}
Solid-solid phase transitions can lead to significant changes of particular material parameters under moderate stimuli. 
Electronic, structural, magnetic and ferroelectric phase transitions and related changes of electronic conductivity, crystal structure, magnetic and ferroelectric ordering offer an opportunity to create functional elements for memories \cite{Wuttig-NMater2007}, neuromorphic computing \cite{Yi-NComm2018}, nanoactuators \cite{Ma-Nanoscale2018}, smart windows \cite{chen2019gate,Huang-APL2012}, field-effect transistors \cite{li2011external, shukla2015steep, ahn2003electric}, sensors \cite{hu2010external}. 
Due to such a diversity of potential applications, identification of stimuli for switching materials between different phases with the highest efficiency in terms of energy consumption and timescale is an essential task.
Clearly, pico- and femtosecond laser pulses in optical, infrared, and terahertz ranges are unmatched stimuli for phase transitions, allowing both switching between the different states at the comparable timescale and for studying dynamics of various medium subsystems during such a conversion. 

The task of revealing involved mechanisms of photo-induced phase transitions is hindered by highly-nonequilibrium nature of the state of matter occurring during and directly after the excitation, and often involving hidden and metastable phases unknown in equilibrium \cite{Wegkamp_PSS2015,Ichikawa-NMater2011,DeJong-nat2013}. 
Furthermore, a spatial inhomogeneity of the excitation and of an emerging new phase also plays a crucial role and affects how the system as a whole is switched between the different phases \cite{li-natcom2022,Dolgikh_ArXiv2022,Ju_PRL2004}.

Among numerous materials with phase transitions, magnetite Fe$_3$O$_4$ occupies a unique place  possessing linked electronic and structural Verwey \cite{Verwey_1939} and spin-reorientation (SR) transitions \cite{bickford1953low,belov-1976spin}, relying on quenching of charge- and orbital ordering as the temperature exceeds a range of 123--130~K \cite{Lorenzo-PRL2008,senn-nat2012}.
It has been already shown that the photo-induced Verwey transition from a monoclinic insulating phase to a cubic metallic phase in Fe$_3$O$_4$ is non-thermal at an early, subpicosecond, stage \cite{DeJong-nat2013}, and the role of laser-induced heating consists in stabilizing the latter phase.
On the other hand, photo-induced SR transition when the magnetic anisotropy changes from uniaxial to cubic and the anisotropy axes reorient by $55^{\circ}$, is still to be demonstrated.
If the link between Verwey and SR transitions still holds for the photo-induced transitions, such a change of magnetic anisotropy would also occur at subpicosecond timescale.
The latter would distinguish photo-induced SR transition in magnetite from those demonstrated in a group of rare-earth orthoferrites \cite{Kimel_Nature2004,deJong-PRL2012,Afanasiev_PRL2016}, with SR transition time exceeding several picoseconds due to slow response of the rare-earth subsystem to the excitation \cite{deJong_PRB2011}.

In this Article we report on the time-resolved optical and magneto-optical study of excitation of a single crystalline magnetite sample with femtosecond near infrared laser pulses in a temperature range of 80--180~K below and above transitions temperature.  
By examining laser fluence and temperature dependence of transient reflectivity and linear polar (PMOKE) and quadratic (QMOKE) magneto-optical Kerr effects, we reveal coupled photo-induced Verwey and SR transitions with the same fluence and temperature thresholds.
Remarkably, photo-induced magnetization precession with the frequency characteristic of a cubic phase, being a signature of the SR transition, is found in the whole studied range of temperatures, as well as upon laser excitation with the fluence below threshold required to heat the sample up to the transition temperature. 
Comparison of PMOKE and QMOKE transients reveals that there is extended range of laser fluences and sample temperatures, where the photo-induced transition occurs in separate domains of material, which highlights the first-order of these phase transitions beyond the range of thermodynamic
equilibrium.

This Article is organized as follows.
In Sec.~\ref{sec:sample} we discuss the main properties of magnetite, Verwey and SR phase transitions, introduce the sample and outline the experimental strategy.
Sec.~\ref{sec:results} presents the experimental data on laser-induced dynamics of reflectivity and magneto-optical effects, which allows identifying photo-induced Verwey and SR transitions.
In Sec.~\ref{sec:discussion} we discuss the experimental results and show how the features of photo-induced SR transition can be extracted from the comparative analysis of the PMOKE and QMOKE transients.
In Conclusions we summarize our findings and suggest an outlook of further development of studies of photo-induced SR transition in magnetite.

\section{Sample and experimental details}\label{sec:sample}

The magnetite Fe$_3$O$_4$ at a room temperature is in the spinel type cubic crystal structure with the number of molecular units in the crystallographic unit cell $Z=8$ \cite{structure}.
Magnetite is a ferrimagnet with a high Curie temperature $T_C=$~860\,K.
A characteristic feature of magnetite is the presence of tetrahedral and octahedral positions in the unit cell, which can both be occupied by Fe$^{3+}$(3d$^5$) and Fe$^{2+}$(3d$^6$) cations.
In fact, the possibility of magnetic ions to occupy both types of positions changes with temperature, thus determining many of the properties of magnetite, and the presence of structural and magnetic phase transitions \cite{Verwey_1939,iizumi1982structure, Lorenzo-PRL2008}.

Below room temperature, Fe$_3$O$_4$ possesses two phase transitions. 
The first-order structural insulator-to-metal transition from monoclinic (space group $P$2$/c$) to cubic (space group $Fd\overline{3}m$,) phase known as a Verwey transition occurs at $T_\mathrm{V}=123$~K \cite{Verwey_1939}. 
An important consequence of this process is an abrupt increase of conductivity by about two orders of magnitude that allows to trace equilibrium and ultrafast Verwey transition, in particular, by measuring static and laser-induced transient reflectivity changes \cite{Sokoloff-PRB1971, rlinger-pss1977, DeJong-nat2013}.

\begin{figure}
\includegraphics [width=0.5\textwidth]  {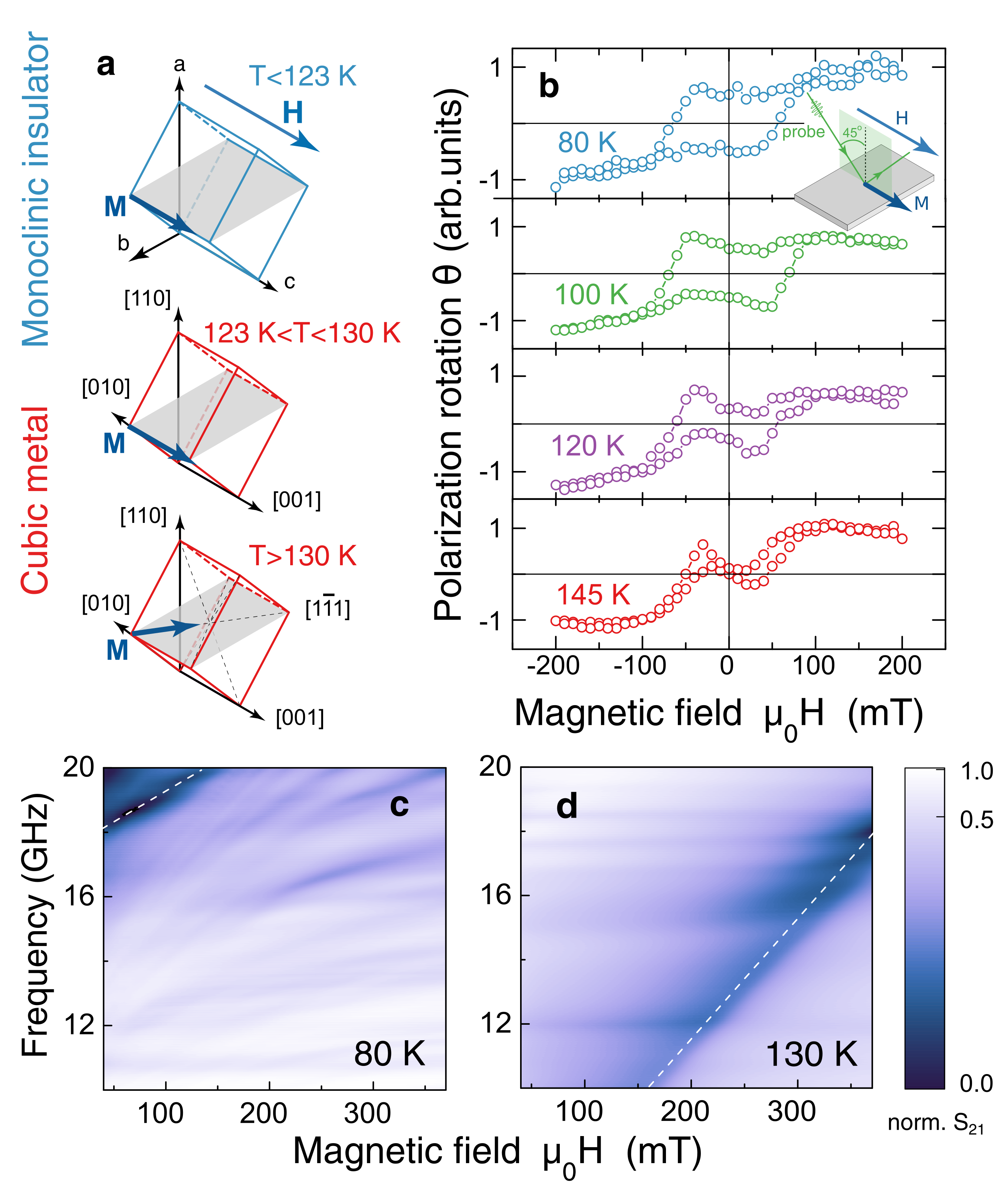}
\caption{\label{fig:Fig1} (a) Crystallographic unit cell and easy magnetization directions below $T_\mathrm{V} = 123$~K (top), at $T = 123-130$~K, and above $T_\mathrm{SR} = 130$~K (bottom).
The sample plane $(110)$ is colored in gray.
In the monoclinic phase the deviation of the crystallographic angles from $90^\circ$ is not shown.
Orientations of the magnetization along easy axis of the corresponding phase are shown. (b) Rotation $\theta$ of probe pulses polarization plane as a function of an external magnetic field measured at different temperatures without pump excitation. \textcolor{revised}{Normalized frequency-field absorption maps at temperatures (c) $T_0=$80~K and (d) 130 ~K. White dashed lines mark the resonant absorption.}} 
\end{figure}

The magnetocrystalline anisotropy is intrinsically linked to the crystal structure symmetry.
Below $T_\mathrm{V}$, in the monoclinic phase, magnetite possesses uniaxial anisotropy with the easy magnetization axis lying along the crystallographic $c$~direction, as shown in Fig.\,\ref{fig:Fig1}~(a).
Above $T_\mathrm{V}$, magnetite is characterized by a cubic magnetocrystalline anisotropy.
Anisotropy energy is expressed as \cite{Calhoun-PR1954}:

\begin{equation} \label{eq:anisotropy_energy}
  \begin{split}
  U_\mathrm{uni}&=K_a \alpha_a^2 + K_b \alpha_b^2 \\
  &\qquad + K_{aa} \alpha_a^4 + K_{bb}\alpha_b^4 + K_{ab} \alpha_a^2\alpha_b^2,\\
  U_\mathrm{c}&=K_1(\alpha_1^2\alpha_2^2+\alpha_2^2\alpha_3^2+\alpha_3^2\alpha_1^2)\\
  &\qquad+K_2(\alpha_1^2\alpha_2^2\alpha_3^2),
  \end{split} \,
  \begin{split}
  \quad\\
  &T<T_V\\
  \quad\\
  &T>T_V
  \end{split}
\end{equation}
where $K_{a,b,aa,bb,ab}$, are the uniaxial, $K_{1, 2}$ are the cubic anisotropy parameters, $\alpha_{a(b)}$ and $\alpha_{1, 2, 3}$ are the directional cosines of the magnetization vector with respect to the monoclinic $a(b)$ and cubic $\langle100\rangle$ axes, respectively. 
In the temperature range of 123--130\,K, parameters $K_1$ and $K_2$ are positive and the easy axes are oriented along the $\langle100\rangle$ crystallographic directions \cite{K1andK2}. 
At $T_\mathrm{SR}=130$\,K, the anisotropy constants change sign that results in a SR transition marked by reorientation of the easy axes to the cube diagonals $\langle111\rangle$ \cite{muxworthy2000review} as illustrated in Fig.\,\ref{fig:Fig1}~(a).
Equilibrium and laser-induced SR transitions can thus be detected by tracing magnetization orientation by means of magneto-optical effects.

Bulk single crystal Fe$_3$O$_4$ was grown by the floating zone method \cite{balbashov1981apparatus}. 
A 1-mm thick plane parallel plate was cut perpendicularly to the crystallographic axis $[110]$ (in the cubic phase) and polished to optical quality. 
This sample orientation was chosen because in the cubic phase the $(110)$ plane contains two $\langle111\rangle$ axes and [001] axis as well as $c$-axis in the monoclinic phase, i.e. the SR transition $[001]\rightarrow\langle111\rangle$ occurs in this plane, given the presence of a demagnetizing field $\mathbf{H}_\mathrm{d}$, as illustrated in Fig.~\ref{fig:Fig1}~(a).

The laser-induced phase transitions in a bulk sample of magnetite were studied using a femtosecond two-color magneto-optical pump-probe technique. 
Pump and probe pulses with duration 170\,fs were emitted by Yb$^{3+}$:KGd(WO$_4$)$_2$ regenerative amplifier at a repetition rate of 100\,kHz.
Pump pulses with a central photon energy 1.2 \,eV were focused normally to the sample surface into an area with a diameter of 80~$\mu$m.
Probe pulses with double photon energy are focused into the spot with a diameter of 25~$\mu$m at an incidence angle of 45$^{\circ}$.
The pump fluence $F$ was varied in a range of 0.7--11~mJ/cm$^2$, exceeding probe fluence by up to 50 times.
Pump-induced reflectivity change $\Delta R$ and magneto-optical rotation of the probe polarization $\Delta \theta$ were measured as a function of pump-probe time delay $\Delta t$ to monitor laser-induced Verwey and SR transitions, respectively.

Experiments were performed at the sample temperature $T_0$ in a range of 80--180~K. The external magnetic field $\mu_0H=$50--250~mT was applied in the sample plane along the cubic $[001]$ crystallographic axis and close to the monoclinic $c-$axis.
Cooling the sample down below $T_\mathrm{SR}$ and $T_\mathrm{V}$ was carried out in an external field $\mu_0H=250$~mT applied along the same direction which suppresses emergence of structural domains in the monoclinic phase with orthogonal orientations of $c-$axes \cite{bickford1953low}.

In order to characterize magnetic anisotropy of the sample at equilibrium, probe polarization rotation $\theta$ was measured at different temperatures $T_0$ as a function of an external magnetic field $H$ without pump excitation [Fig.~\ref{fig:Fig1}~(b)].
In our experimental geometry with equilibrium in-plane orientation of magnetization, $\theta$ was determined by the magnetization component along the field via longitudinal magneto-optical Kerr effect \cite{zvezdin-book1997}.
Hysteresis loops measured at $T_0=80$~K and 145~K correspond to the magnetization easy-axis directed along $c$-axis in monoclinic phase and along $[1\bar11]$ in cubic phase, respectively [Fig.~\ref{fig:Fig1}~(b)].
Noticeably, at intermediate temperatures the dependence $\theta$ on $\mu_0H$ represents the superposition of these two types of hysteresises.
This indicates the coexistence of phases with uniaxial and cubic anisotropies in a wide temperature range in agreement with the first-order of phase transitions 
\cite{NewDomains}.

\textcolor{revised}{ For further characterization, ferromagnetic resonance (FMR) absorption measurements were performed using the vector-network analyzer method \cite{kalarickal2006ferromagnetic, maksymov2015broadband, neudecker2006comparison}. 
Absorption maps were obtained as a series of frequency sweeps measured at fixed magnetic field.
The sample was placed on the coplanar waveguide attached to the cold finger of the cryostat. 
The response of the system is studied by analyzing the transmitted microwave signal S$_{21}$ [Fig.\ref{fig:Fig1}~(c,~d)].
The signal from the sample was normalized on the response from the waveguide.
The external magnetic field was applied as described above, i.e. in the sample plane along the cubic [001] crystallographic axis.}

\section{Results}\label{sec:results}
\begin{figure}
\includegraphics [width=0.46\textwidth] {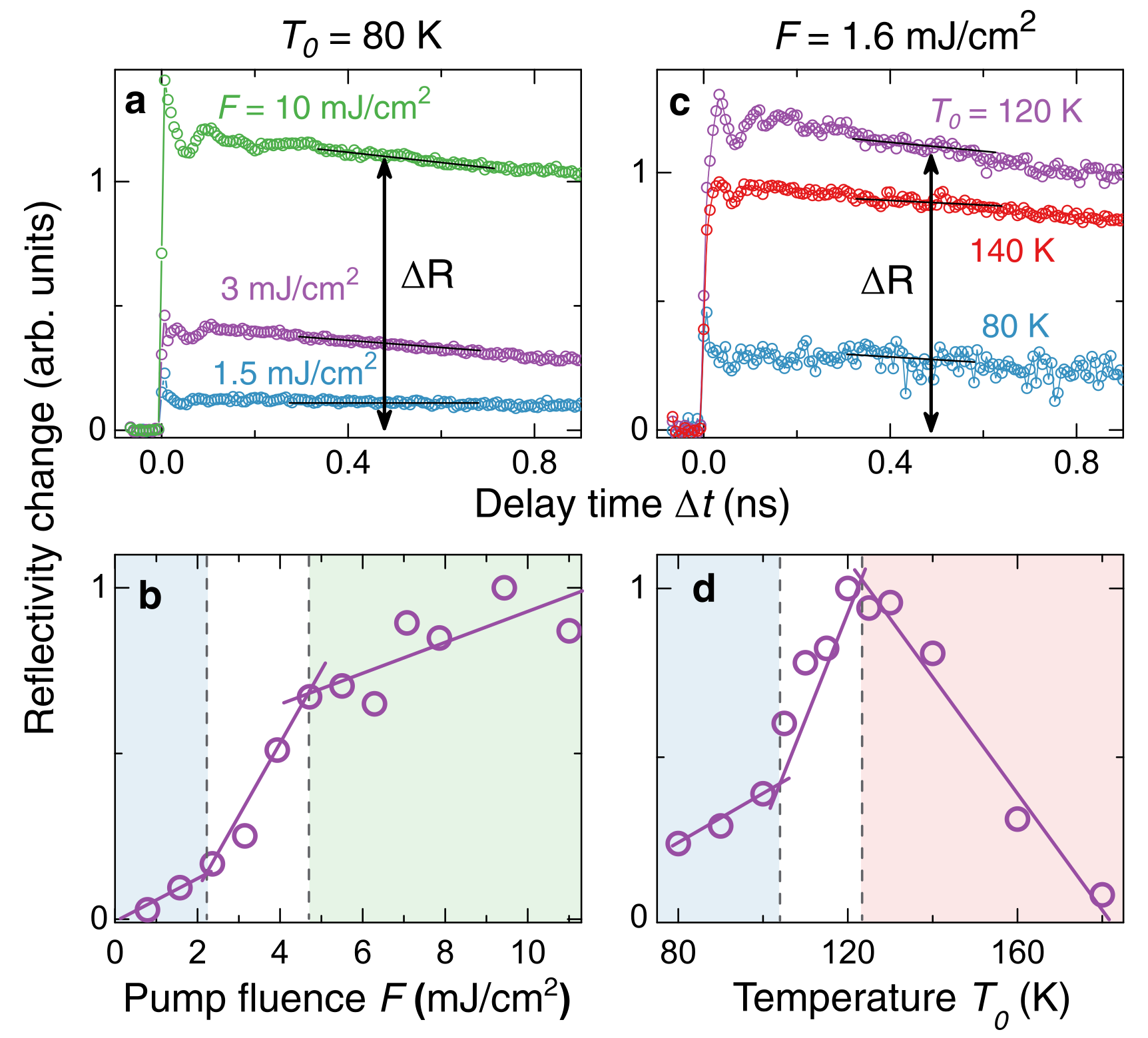}
\caption{\label{fig:dR_vs_temp_flu} Reflectivity change $\Delta R$ as a function of the time delay $\Delta t$ between pump and probe pulses (a) at fixed initial temperature $T_0=$80\,K for different values of the pump fluence $F$ and (c) at a fixed pump fluence $F=$1.6\,mJ/cm$^2$ for different $T_0$.
Laser-induced reflectivity change $\Delta R$ at $\Delta t=0.5$~ns as a function of (b) pump fluence $F$ and (d) $T_\mathrm{0}$. 
Solid lines on panels (b,d) are linear fit. Calculated laser-induced heating temperature of the sample $T_h$ as a function of (b) the pump fluence at fixed initial temperature $T_0$ = 80 K and (d) initial temperature at $F=$1.6\,mJ/cm$^2$ are shown as pink solid lines.}
\end{figure}

First we verify that the Verwey transition is induced in the sample by the pump pulses.
Figure~\ref{fig:dR_vs_temp_flu}~(a) shows typical laser-induced reflectivity change $\Delta  R$ measured as a function of pump-probe time delay $\Delta t$ for three values of pump fluence $F$ at the initial sample temperature $T_0=80$~K well below $T_{V}$. 
Abrupt change of reflectivity followed by slow relaxation is typical for the ultrafast Verwey transition \cite{DeJong-nat2013, lysenko2006light}.
Laser-induced change $\Delta  R$ at $\Delta t=0.5$~ns is plotted as a function of pump fluence $F$ in Fig.~\ref{fig:dR_vs_temp_flu}\,(b).
There are two features marked with vertical dashed lines, at which the slope of $\Delta R(F)$ changes.
Such a nonlinear behaviour of $\Delta R(F)$ is a clear indication of the laser-induced first-order insulator-to-metal transition and typically interpreted as follows.

At fluences below the threshold $F_\mathrm{th}=2.2$~mJ/cm$^2$, the transient reflectivity is dominated by the optical response of the material which remains in a low-temperature phase.
In a range between $F_\mathrm{th}$ and saturation fluence $F_\mathrm{S}=4.7$~mJ/cm$^2$, the superlinear increase indicates that the increasing fraction of the excited material switches to the cubic metallic phase and gives rise to the reflectivity change \cite{fe3o4_Ref_broadband}. 
Above $F_\mathrm{S}$, the response is dominated by material which experienced laser-induced phase transition.
We note that the threshold and saturation fluences are in a good agreement with the literature data \cite{pontius2011time}.

Also, laser-induced reflectivity change $\Delta R$ at various initial sample temperatures $T_0$ was measured with the fixed pump fluence $F=1.6$~mJ/cm$^2$ [Fig.~\ref{fig:dR_vs_temp_flu}\,(c)].
Temperature dependence of $\Delta R$ demonstrates two features as well [Fig.~\ref{fig:dR_vs_temp_flu}~(d)]. 
Increase of the slope of $\Delta R(T_0)$ at $T_\mathrm{th}\approx100$~K suggests that starting from this temperature fraction of material, switched to the metallic phase, starts to contribute to the transient reflectivity signal, i.e. $F_\mathrm{th}$ decreases down to 1.6~mJ/cm$^2$ at this temperature.
The second feature coincides with $T_\mathrm{V}=123$~K.
As $T_0$ exceeds $T_\mathrm{V}$ the slope of $\Delta R(T_0)$ becomes negative, and the underlying picture of this is discussed below.

\begin{figure}
\includegraphics [width=0.46\textwidth] {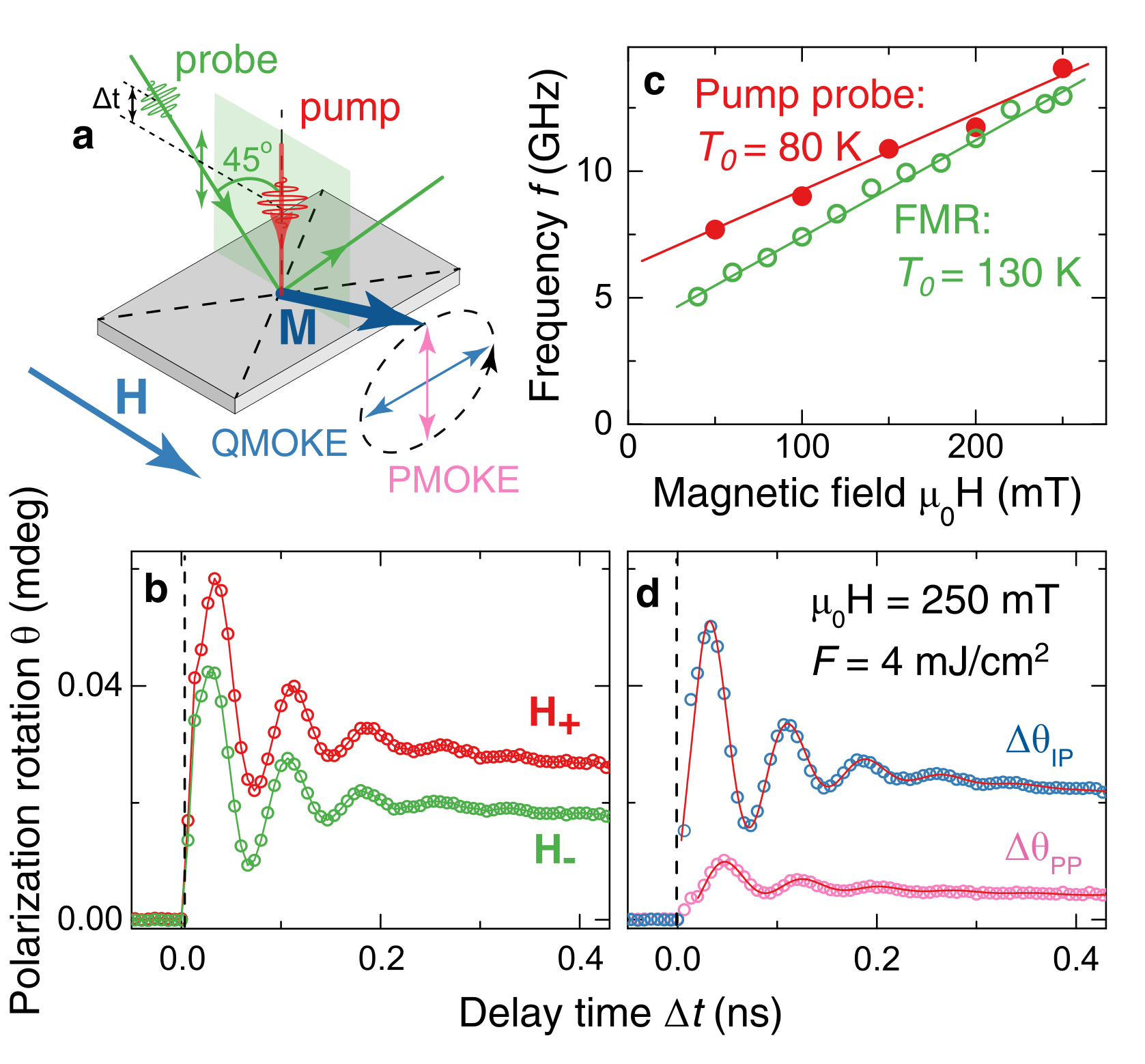}
\caption{\label{fig:precession_theta_PPIP_extract} 
(a) Scheme of experimental geometry.
An external magnetic field $\mathbf{H}$ is applied along $c([001])$ direction.
Two fold arrows stand for magnetization transient changes revealed by PMOKE and QMOKE.
(b) Laser-induced probe polarization rotation $\Delta \theta$ as a function of time delay $\Delta t$ measured in the magnetic field $\mu_0H=\pm250$~mT. 
(c) Oscillation frequency $f$ (closed symbols) as a function of magnetic field $\mu_0H$ at $T_0=80$\,K, $F=4$\,mJ/cm$^2$ and $f_\mathrm{FMR}$ (open symbols) as a function of magnetic field $\mu_0H$ at $T_0=130$\,K.(d) Extracted signals $\Delta \theta_{PP}(\Delta t)$ and $\Delta \theta_{IP}(\Delta t)$ (see text for details). }
\end{figure}

In order to demonstrate ultrafast pump-induced SR transition in magnetite we measured transient magneto-optical response 
to the laser excitation.
Typical signals $\Delta \theta(H_{\pm};\Delta t)$ measured in the external field $\mu_0H_\pm=\pm250$~mT are shown in Fig.~\ref{fig:precession_theta_PPIP_extract}~(b).
Field dependence $f(\mu_0H)$ of the frequency of the observed oscillations indicates that there is a magnetization precession triggered by pump pulses [Fig.~\ref{fig:precession_theta_PPIP_extract}~(c)].
In our experimental geometry, measured probe polarization change is sensitive to oscillations of both in-plane (IP) and perpendicular-to-plane (PP) magnetization components via 
QMOKE and PMOKE, respectively, as illustrated in Fig.~\ref{fig:precession_theta_PPIP_extract}~(a).
We note that linear and quadratic MOKE in a ferrimagnet Fe$_3$O$_4$ can be comparable, as was indeed reported for the probe photon energy \cite{silber2018quadratic}.

Taking into account that QMOKE and PMOKE are even and odd effects with respect to magnetization reversal, respectively, we distinguish contributions from oscillations of the IP ($\Delta \theta_{IP}$) and PP ($\Delta \theta_{PP}$) components of magnetization to the measured rotation as $\Delta \theta_{IP(PP)}(\Delta t)=0.5(\Delta \theta(H_{+};\Delta t)\pm\Delta \theta(H_{-};\Delta t))$.
Thus obtained $\Delta \theta_{IP(PP)}(\Delta t)$ both show oscillatory behaviour and are fitted with the exponentially damped sine-function [Fig.~\ref{fig:precession_theta_PPIP_extract}~(d)]:
\begin{equation}\label{eq:fit}
\Delta\theta_{j}(\Delta t)=\Theta_{j}\exp(-\Delta t/\tau_j)\sin\left(2\pi f\Delta t+\phi_{j}\right),
\end{equation}
where $\Theta_{j}$, $f$, $\tau_j$ and $\phi_{j}$ are the amplitude, frequency, damping time and initial phase of the oscillations, respectively, $j$ stands for PP, IP.

\begin{figure}
\includegraphics [width=\columnwidth]  {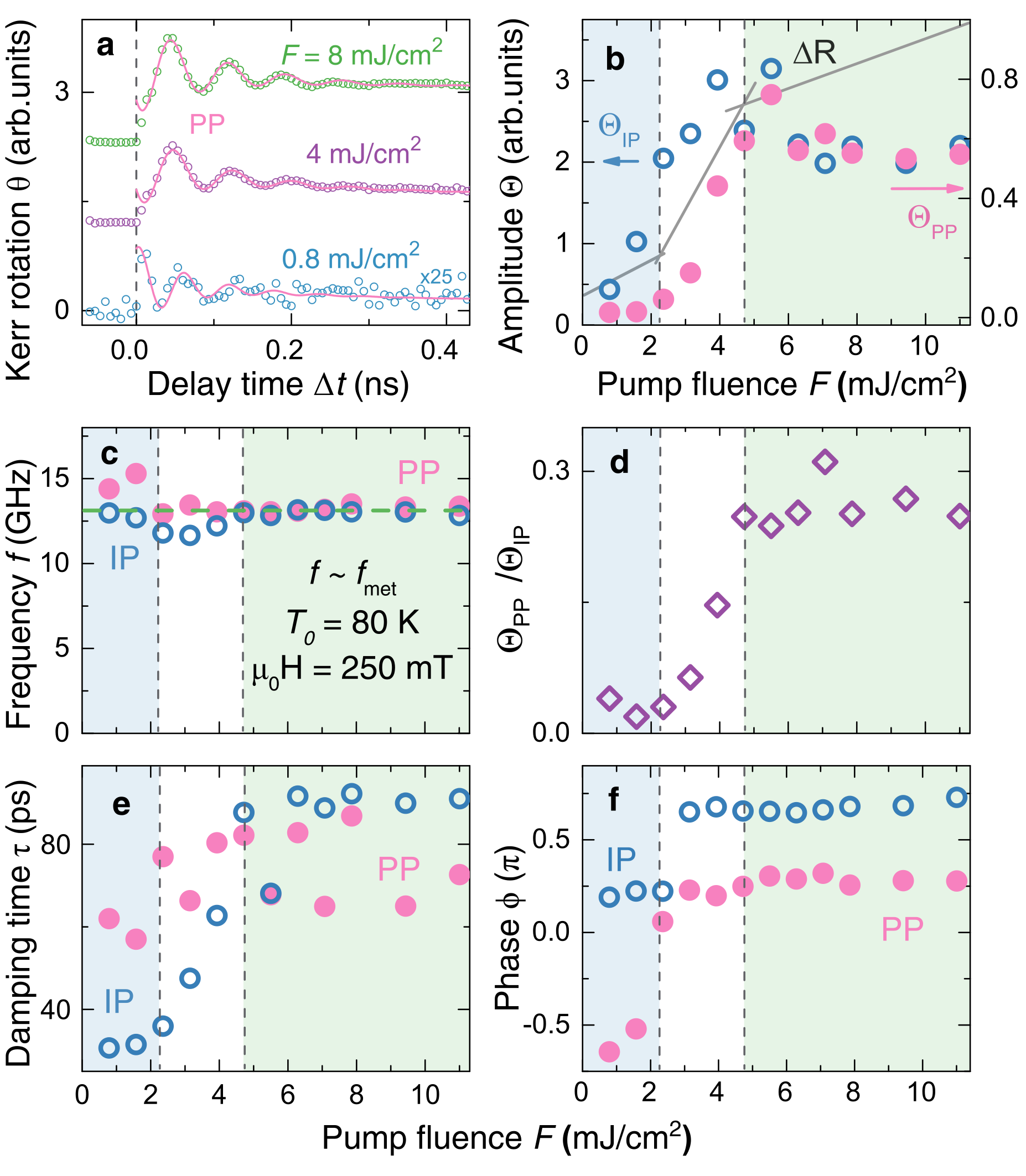}
\caption{\label{fig:Prec_param_VS_flu}  (a) Typical pump-probe signals $\theta_{PP}(\Delta t)$ (symbols) measured at $T_0=80$~K at the pump fluences $F<F_\mathrm{th}$ (blue), $F_\mathrm{th}<F<F_\mathrm{S}$ (magenta), and $F>F_\mathrm{S}$ (green), and their fit (lines) using Eq.~(\ref{eq:fit}). Pump fluence dependences of (b) amplitude $\Theta_{PP(IP)}$, (c) frequency $f_{PP(IP)}$, (d) ratio $\Theta_{PP}/\Theta_{IP}$, (e) damping time $\tau_{PP(IP)}$, and initial phase $\phi_{PP(IP)}$ at $T_0=80$\,K obtained from the fit of $\theta_{PP(IP)}(\Delta t)$ traces.
Closed (open) symbols show the parameters of $\theta_{PP(IP)}(\Delta t)$, respectively.
Gray line in (b) is the fit of $\Delta R(F)$ dependence [Fig.~\ref{fig:dR_vs_temp_flu}(b)].
Green dashed line in (b) shows the FMR frequency measured at $T=$130~K.}
\end{figure}

In Fig.\,\ref{fig:Prec_param_VS_flu} parameters of oscillations $\Delta \theta_{PP}$ and $\Delta \theta_{IP}$ measured at $T_0=80$~K are plotted as a function of pump fluence $F$ in a range of 0.7--11~mJ/cm$^2$.
Note, that in the whole range of fluences, frequencies $f_{PP}\approx f_{IP}$ [Fig.~\ref{fig:Prec_param_VS_flu}~(c)], while initial phases $\varphi_{PP(IP)}$ differ by $\sim\pi/2$ [Fig.~\ref{fig:Prec_param_VS_flu}~(f)], thus confirming that $\Delta \theta_{PP}$ and $\Delta \theta_{IP}$ stem from oscillations of the PP and IP components of magnetization [Fig.\,\ref{fig:precession_theta_PPIP_extract}~(a,~d)].

Importantly, there is a qualitative correspondence between fluence dependences of the reflectivity transient change $\Delta R(F)$ and of some of the oscillations parameters.
To highlight it, we plot in Fig.~\ref{fig:Prec_param_VS_flu}(b) the fitting curve of $\Delta R(F)$ alongside with the oscillations amplitude, and show corresponding ranges $F<F_\mathrm{th}$ and  $F>F_\mathrm{S}$ in the fluence dependences of all the parameters in Fig.~\ref{fig:Prec_param_VS_flu}.
First, when pump fluence exceeds $F_\mathrm{S}$, amplitudes $\Theta_{PP(IP)}$, 
their ratio $\Theta_{PP}/\Theta_{IP}$, damping time $\tau_{PP(IP)}$ and frequency $f$ are independent from $F$.
In the range $F_\mathrm{th}<F<F_\mathrm{S}$ there is a linear growth of $\Theta_{PP}$, similarly to $\Delta R(F)$, whereas the precession frequency $f$ changes only slightly.
Such behaviour agrees in general with the laser-induced SR transition. 
When pump fluence $F$ becomes sufficient for SR transition, magnetization precession is triggered due to switching of magnetic anisotropy axis from the $c$-axis to the cubic $\langle111\rangle$ direction. 
Herewith, increase and saturation of the amplitude $\Theta_{PP}$ is associated with a gradual increase of the probed material fraction where SR transition occurred until it reaches 100~\%.

However, in the range of pump fluence between $F_\mathrm{th}$ and $F_\mathrm{S}$ there is also a growth of $\Theta_{PP}/\Theta_{IP}$ [Fig.~\ref{fig:Prec_param_VS_flu}~(d)].
This is related to the fact that $\Theta_{IP}$ and $\Theta_{PP}$ vary with pump fluence differently [Fig.~\ref{fig:Prec_param_VS_flu}~(b)].
Even more intriguing result is seen at $F<F_\mathrm{th}$ when reflectivity data are dominated by the optical response of the insulating phase [Fig.~\ref{fig:dR_vs_temp_flu}~(b)]. 
$\Theta_{IP}$ grows much faster with $F$ than $\Theta_{PP}$, and the dependence $\Theta_{IP}(F)$ can be extrapolated down to $F=0$ [Fig.~\ref{fig:Prec_param_VS_flu}~(b)], showing no threshold behaviour.
Furthermore, the precession frequency in this range remains the same as at the higher fluences showing no features when $F$ is swept through $F_\mathrm{th}$, while initial phase $\phi_{PP(IP)}$ and the damping time $\tau_{PP(IP)}$ significantly change [Fig.~\ref{fig:Prec_param_VS_flu}~(f,~e)].

We have also obtained pump-probe signals $\Delta \theta_{PP(IP)}(\Delta t)$ at different $T_0$ with a fixed fluence $F=1.6~\mathrm{mJ/cm}^2$ which is below $F_\mathrm{th}$ at $T_0=80$~K. Fig.~\ref{fig:Prec_param_VS_T0} summarizes oscillation parameters as a function of $T_0$. 
Similarly to the fluence dependence, there is nontrivial behaviour of the amplitudes, damping times and initial phases below $T_\mathrm{th}$, while $f_{PP(IP)}$ varies only weakly.

These observations naturally rise a question on the origin of the precession excited by pump pulses with fluence below $F_{\mathrm{th}}$ at low temperatures, on the interpretation of results at $F>F_{\mathrm{th}}$ as the manifestation of the laser-induced SR transition, as well as on the mechanism undelying the precession excitation at $T_0>T_\mathrm{SR}$.

\section{Discussion}\label{sec:discussion}

\subsection{Precession frequency as a fingerprint \\of a laser-induced SR transition}
Experimental results of laser-induced reflectivity change, in particular, presence of the threshold and saturation fluences, suggest that the laser-induced heating is responsible for stabilizing the laser-induced metallic cubic phase \cite{DeJong-nat2013, park-PRB1998}.
Using literature data on latent heat \cite{latent_heat} and specific heat capacities for metallic and insulator phases of magnetite \cite{heat_capacities}, the temperature increase $\Delta T$ and the resulting temperature $T_\mathrm{h}$ were calculated \cite{Mogunov_NatureComm2020} as a function of $F$ at $T_0=80$~K (See App.~\ref{app:heating_calculation} and Fig.~\ref{fig:Apendix}~(a) therein).
According to that, pump fluence exceeding $F_\mathrm{S}=4.7~\mathrm{mJ/cm^2}$ is sufficient for stabilizing cubic metallic phase emerged because of laser-induced Verwey transition. 
Slow relaxation of $\Delta R(\Delta t)$ indicates that the sample remains in the laser-induced phase longer than 1~ns.
Thus, we can use the frequency of the excited precession as a fingerprint for identifying particular magnetic phase state of the sample within this time range after the excitation.
 
\begin{figure}
\includegraphics [width=0.5\textwidth]  {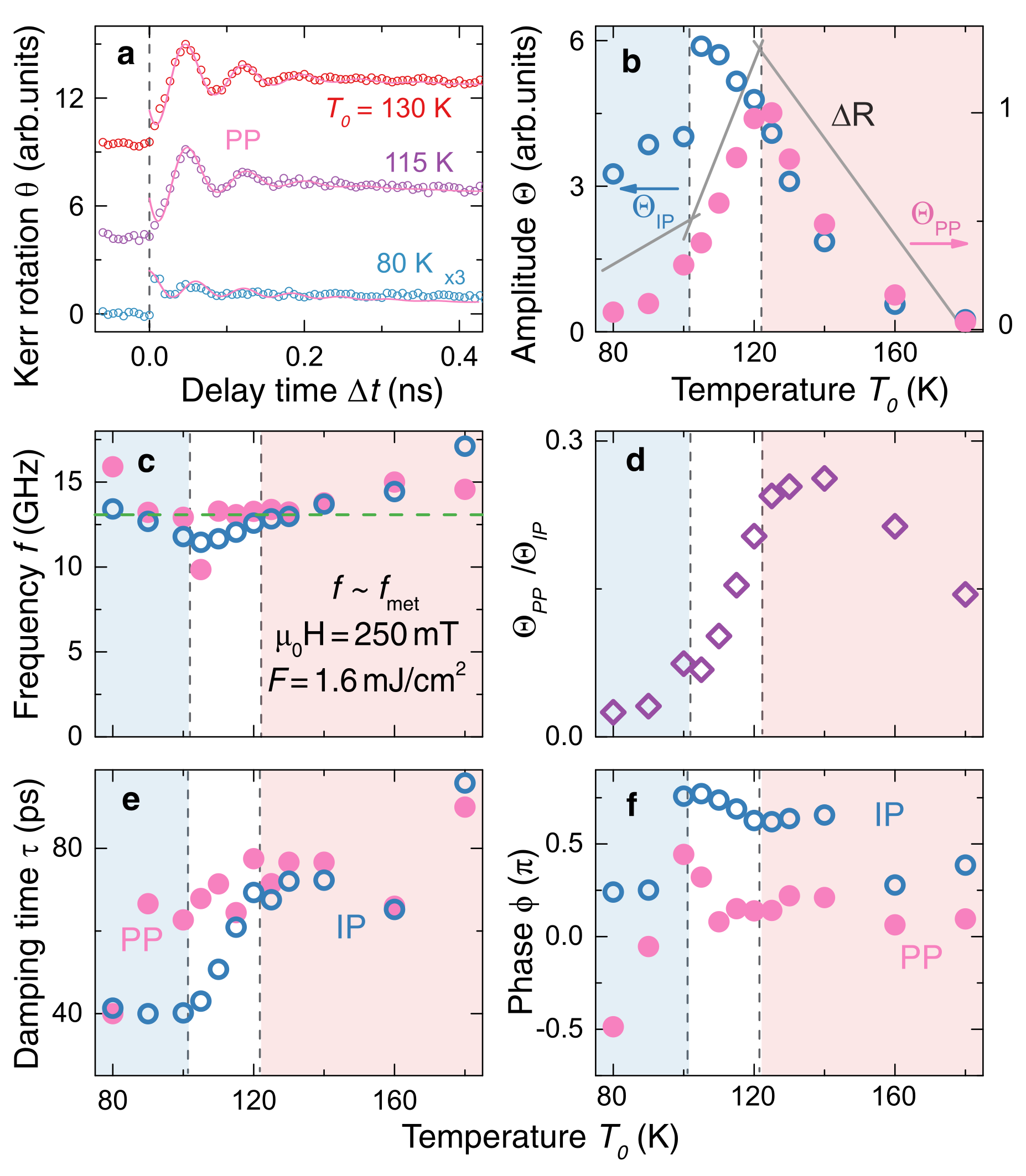}
\caption{\label{fig:Prec_param_VS_T0} 
(a) Experimental pump-probe signals $\theta_{PP}(\Delta t)$ (symbols) measured at $F=1.6$~mJ/cm$^2$ at the sample temperature $T_0<T_\mathrm{th}$ (blue), $T_\mathrm{th}<T_0<T_\mathrm{V}$ (magenta), and $T_0>T_\mathrm{V}$ (red), and their fit (lines) using Eq.~(\ref{eq:fit}). 
Temperature dependences of (b) amplitude $\Theta_{PP(IP)}$, (c) frequency $f_{PP(IP)}$, (d) ratio $\Theta_{PP}/\Theta_{IP}$, (e) damping time $\tau_{PP(IP)}$, and initial phase $\phi_{PP(IP)}$ at $F=1.6$~mJ/cm$^2$ obtained from the fit of $\theta_{PP(IP)}(\Delta t)$ traces.
Closed (open) symbols show the parameters of $\theta_{PP(IP)}(\Delta t)$, respectively.
Gray line in (b) is the fit of $\Delta R(T_0)$ dependence [Fig.~\ref{fig:dR_vs_temp_flu}(b)].
Green dashed line in (b) shows the FMR frequency measured at $T=$130~K.}
\end{figure}

In order to illustrate this, we show in Fig.\ref{fig:Fig1}~(c,~d) FMR frequency $f_\mathrm{FMR}$ as a function of external magnetic field $H$ measured at equilibrium at $T_0=80$ and $130$~K.
As shown in Fig.\,\ref{fig:Prec_param_VS_flu}~(c) and Fig.\,\ref{fig:Prec_param_VS_T0}~(c), precession frequency in the pump-probe experiment in the whole fluence and temperature ranges is close to the $f_\mathrm{FMR}$ above $T_\mathrm{SR}$, i.e. in the cubic phase. 
Note, the frequency in the monoclinic phase is expected to be twice as high, as was calculated using anisotropy parameters from \cite{K1andK2,abe_Ka} 
(See App.~\ref{app:FMR_calculation} and Fig.~\ref{fig:Apendix}~(c) therein). 
At $\mu_0H=250$~mT and $T_0=80$~K, it exceeds $20$~GHz available in the FMR experiments [Fig.~\ref{fig:Fig1}~(c)].
Thus, the frequency of the observed precession clearly indicates that SR transition takes place even when the sample is excited at $T_0$=80~K with low-fluence pulses insufficient for heating above the transitions temperatures.
On the other hand, experimental data on the ratio $\Theta_{PP}/\Theta_{IP}$, the decay time $\tau_{PP(IP)}$ and the initial phases $\phi_{PP(IP)}$ [Figs.~\ref{fig:Prec_param_VS_flu}~(d-f)] reveal that there is a difference between precession excited at $F>F_\mathrm{S}$ and $F<F_\mathrm{th}$, despite similar precession frequency observed in the whole laser fluence range.

\textcolor{revised}{Absence of the laser-induced precession with frequency corresponding to the low-temperature uniaxial state can be understood from the following arguments.
The direction of the easy magnetization axis in the low-temperature phase coincides with the applied external field.
This along with a strong uniaxial anisotropy ensures that no magnetization precession is excited by the laser pulses unless the SR transition is induced.
Thus, presence of the oscillations in the MOKE signal suggests that precession is excited only in the fraction of the material undergoing laser-induced SR transition.}

\subsection{Three regimes of SR transition \\at various pump fluences}

We argue that our experimental observations are the manifestation of peculiarities of laser-induced SR transition being the first-order at equilibrium.
At $T_0=80$~K and $F>F_\mathrm{S}$, the precession excitation can be described as follows.
Before the excitation, magnetization orientation is defined by the effective field $\mathbf{H}_\mathrm{eff}=\mathbf{H}+\mathbf{H}_\mathrm{a}+\mathbf{H}_\mathrm{d}$ directed close to the $c-$axis, where $\mathbf{H}_\mathrm{a}  = -\partial U_\mathrm{uni}(K_a, K_b)/\partial \mathbf{M}$ is the uniaxial anisotropy field.
Laser pulse causes ultrafast Verwey transition and, thus, switching of the crystal structure to the cubic phase, which is stabilized by the laser-induced heating above $T_\mathrm{V}$ \cite{DeJong-nat2013,Pontius-APL2011}. 
Structural transition is accompanied by the change of magnetic anisotropy to the cubic one.
Since $T_\mathrm{SR}$ is just a few Kelvin above $T_\mathrm{V}$, $K_{1,2}$ change sign,
 and the anisotropy field reorients to one of 
$\langle111\rangle$ axes lying in the sample plane, e.g. to $[1\bar11]$. 
As a result, there is a PP torque $\mathbf{T}_{PP} = -\gamma \mathbf{M}\times\mathbf{H'_\mathrm{eff}}$.
The torque triggers the precession of magnetization around $\mathbf{H'_\mathrm{eff}}=\mathbf{H}+\mathbf{H'}_\mathrm{a}+\mathbf{H}_\mathrm{d}$, where $\mathbf{H'}_\mathrm{a}=-\partial U_\mathrm{c}(K_1, K_2)/\partial\mathbf{M}$. Initial phases of the oscillations of the PP and IP components are $0$ and $\pi/2$, respectively, in agreement with experimental observations [Fig.~\ref{Fig:schematics}~(a)].

Even at $T_0<T_\mathrm{SR}$ stable domains with cubic magnetic anisotropy are present within host with uniaxial anisotropy, as evident from the static hysteresis loops [Fig.~\ref{fig:Fig1}~(b)].
This suggests that the below-threshold excitation  can still yield emergence of the domains of a cubic phase within the monoclinic host although laser-induced heating is $T_0+\Delta T(F)<T_\mathrm{th}$.
These domains are expected to be isolated from each other, and thus possess the easy magnetization axes along either of the four cubic axes $\langle111\rangle$ in the sample plane or at an angle to it.
Thus, laser-induced precession in the different domains is then triggered with a different initial phase: 
\begin{equation}
  \begin{pmatrix}
   M_{PP,i} (\Delta t)\\
   M_{IP,i} (\Delta t)\\
  \end{pmatrix} =
 \begin{pmatrix}
   \varepsilon M_0\sin(2\pi f\Delta t+N\frac{\pi}{2})\\
   M_0\sin(2\pi f\Delta t+(N+1)\frac{\pi}{2})\\
  \end{pmatrix},
\end{equation}
where $\varepsilon$ is the precession ellipticity, $i$ stands for a particular domain, and $M_0$ is the precession amplitude given by an angle by which total effective field reorients as a result of SR transition.
Integer number $N$ depends on the particular cubic axis which sets $\mathbf{H}'_\mathrm{a}$.
$N=0,2 (1,3)$ for domains with cubic axis oriented in the sample plane (at an angle to it) with the corresponding laser-induced torque $\mathbf{T}_{PP}$($\mathbf{T}_{IP}$) as illustrated in Fig.~\ref{Fig:schematics}~(a,~b).

In the experiment, measured signal $\theta_{PP(IP)}$ is integrated over the large area of the sample. 
Precession with different initial phases in different domains contribute to $\theta_{PP}\sim\sum_i M_{PP,i}$ yielding very low value of $\Theta_{PP}$ and intermediate initial phase of oscillations $\phi_{PP}\neq0,\pi/2$.
Contributions of the precession in different domains to $\theta_{IP}\sim\sum_i M_{IP,i}^2$ add up resulting in non-vanishing value of $\Theta_{IP}>\Theta_{PP}$ observed even at low values of $F$ [Fig.~\ref{fig:Prec_param_VS_flu}~(a)] and to the intermediate initial phase of oscillations.
Such averaging also explains faster decay at $F<F_\mathrm{th}$ being a result of dephasing of precession in different domains and of the enhanced damping within them \cite{jozsa2006optical, Barman-PRB2009}.

Onset of increase of $\tau$ and $\Theta_{PP}/\Theta_{IP}$ at the fluence $F>F_\mathrm{th}$ [Fig.~\ref{fig:Prec_param_VS_flu}~(d,~e)] and the change of $\phi_{PP(IP)}$ to $\sim0(\pi/2)$ [Fig.~\ref{fig:Prec_param_VS_flu}~(f)] suggests that ever increasing areas and networks where SR transition occurs start to contribute to the magneto-optical response instead of isolated domains  \cite{Hilton_PRL2007, Cocker-PRB2012}.
This leads to an increase of a precession damping time $\tau$ in the range $F_\mathrm{th}<F<F_\mathrm{S}$. 
Increase of $\Theta_{PP}$ along with the switching of the $\phi_{PP(IP)}$ close to $0~(\pi/2)$ shows that the magnetization precession within the probed spot is mostly in-phase and the laser-induced magnetic anisotropy axis is along one particular $\langle111\rangle$ axis in the sample plane [Fig.~\ref{Fig:schematics}~(a)].
At any of the studied $F>F_\mathrm{S}$, the whole excited and probed volume experiences laser-induced SR transition, leading to saturation of the parameters of $\theta_{PP(IP)}$ transients. 
$\Theta_{PP}/\Theta_{IP}$ is then defined by the precession ellipticity $\varepsilon$ and the magneto-optical parameters responsible for PMOKE and QMOKE.

\subsection{Three regimes of SR transition \\at various sample temperatures}

Joint analysis of the evolution of the transients parameters with the initial temperature $T_0$ enables to identify three regimes of laser-induced dynamics.   
At the fixed laser fluence $F=1.6$~mJ/cm$^2$, temperature ranges $T_0<T_\mathrm{th}$ and $T_\mathrm{th}<T_0<T_\mathrm{SR}$ correspond to the same regimes of laser-induced SR transition, as those seen at the fixed $T_0$ at $F<F_\mathrm{th}$ and $F_\mathrm{th}<F<F_\mathrm{S}$, respectively.

Interestingly, our data suggest that even at $T_0>T_\mathrm{SR}$ the precession is excited due to laser-induced SR transition. 
\textcolor{revised}{Indeed, along with the laser-induced SR transition, precession can be triggered by the laser-induced thermal change of magnetic anisotropy parameters while the material remains in the same phase \cite{Shelukhin-PRB2018}.
Below $T_\mathrm{V}$, laser-induced changes of the uniaxial anisotropy parameters as a mechanism of the precession excitation can be ruled out, since the precession frequency corresponds to the cubic phase (see Sec.~\ref{sec:results}).
However, the question arises on the mechanism of the precession excitation at $T_0>T_\mathrm{SR}$.
We ascribe it to the SR transition in remaining metastable monoclinic domains.
For this we put forward the following arguments.}

\begin{figure}
\includegraphics [width=\columnwidth] {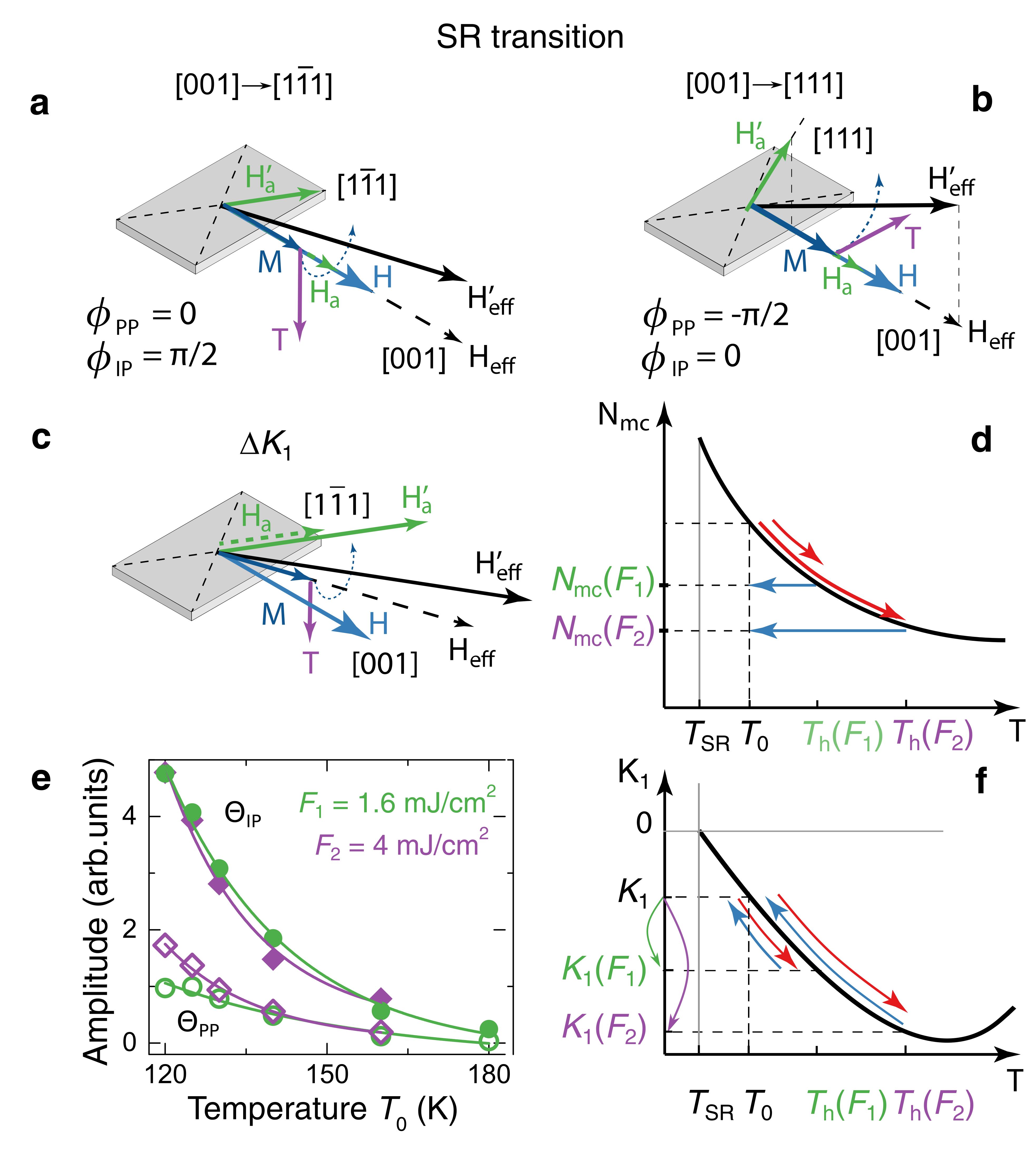}
\caption{\textcolor{revised}{Scheme of precession launch via (a,~b) the SR transition and the easy axis reorientation to different $\langle111\rangle$ directions, and (c) the laser-induced cubic anisotropy change.
(d) Qualitative dependence of the metastable monoclinic domains amount $N_{mc}$ on the  temperature.
(e) Amplitudes $\Theta_{PP}$ (open symbols) and $\Theta_{IP}$ (solid symbols) at the pump pulse fluences $F = 1.6$~mJ/cm$^2$ (blue symbols) and 4~mJ/cm$^2$ (red symbols).
Lines are the exponential fits.
Red and blue arrows indicate heating-cooling cycle after single laser pulse excitation for lower ($F_1$) and higher ($F_2$) fluences.
$N_\mathrm{mc}$ decreases following each heating-cooling cycle.
(f) Dependence of the cubic magnetic anisotropy parameter $K_1$ on the temperature \cite{abe_Ka}.
$K_1$ restores its equilibrium value $K_1(T_0)$ following each heating-cooling cycle.}
}\label{Fig:schematics}
\end{figure}

\textcolor{revised}{In the range of $T_\mathrm{SR}<T_0\lessapprox230$~K absolute value of the magnetocrystalline constant $K_1$ increases with temperature, as illustrated in Fig.~\ref{Fig:schematics}~(f) following \cite{abe_Ka}. 
Since laser-induced heating does not exceed 40~K for any fluence $F$ used in the experiment (see App.\,\ref{app:heating_calculation}), increase of $K_1$ upon laser excitation is the case for measurements in the whole range of initial temperatures.
As a result, reorientation of $\mathbf{H}_\mathrm{eff}$ due to the laser-induced change of $K_1$ and SR transition would occur in the same way, as shown in Fig.~\ref{Fig:schematics}~(a,~c), leading to the same phase of precession excited by these two mechanisms.
However, a simple geometrical analysis shows that the temperature dependences of the amplitude of the precession excited via these two mechanisms are expected to be drastically different.
Indeed, precession amplitude is set by the angle between direction of effective field $\mathbf{H}_\mathrm{eff}$ before and after excitation (Fig.\,\ref{Fig:schematics})}.
\textcolor{revised}{At any of studied $T_0>T_\mathrm{SR}$, $\Delta T\approx$~const [see App.~\ref{app:heating_calculation} and Fig.\,\ref{fig:Apendix}~(b)].
Due to nearly linear dependence of $K_1$ on temperature in the range $T_\mathrm{SR}<T_0\lessapprox230$~K [Fig.~\ref{Fig:schematics}~(f)], the angle between $\mathbf{H'}_\mathrm{a}$ and $\mathbf{H}_\mathrm{a}$ 
should also be nearly independent from $T_0$.
Thus, the amplitude of precession excited above $T_\mathrm{SR}$ via laser-induced change of $K_1$ should vary with temperature only weakly.}

\textcolor{revised}{On the other hand, the amplitude of the observed oscillations triggered by SR transition $T_0>T_\mathrm{SR}$ is dictated by the amount of metastable domains remaining in the monoclinic phase $N_\mathrm{mc}(T_0)$, in which transition can occur.
$N_\mathrm{mc}$ decreases with temperature leading to reduction of the precession amplitude excited due to the SR transition.
In Fig.~\ref{fig:Prec_param_VS_T0}~(b) one can see that precession amplitude significantly decreases with temperature at $T_0>T_\mathrm{SR}$ indicating that underlying mechanism of excitation is the laser-induced SR transition in these metastable domains.
Such scenario accounts well also for the decrease of $\Delta R$ with temperature in the same range [Fig.~\ref{fig:dR_vs_temp_flu}~(d)].}

\textcolor{revised}{Further confirmation of the SR transition responsible for the precession observed at $T>T_\mathrm{SR}$ is provided by the temperature dependences of $\Theta_{PP(IP)}$ obtained at low $F_1= 1.6$~mJ/cm$^2$ and high $F_2= 4$~mJ/cm$^2$ pump fluence.
As can be seen in Fig.\ref{Fig:schematics}~(e), both amplitudes show exponential decrease, which is faster at higher excitation fluence. 
Excitation with a pulse of the higher fluence would affect the amount of monoclinic metastable domains $N_\mathrm{mc}$ available for SR transition as follows.
Each excitation event leads to a heating-cooling cycle $T_0 \rightarrow T_\mathrm{h} \rightarrow T_0$, with $T_0>T_\mathrm{SR}$. At the heating stage, SR transition occurs in some domain. 
At the cooling stage, they remain in cube phase.
As a result of such a cycle, the number of monoclinic domains $N_\mathrm{mc}$ would decrease [Fig.~\ref{Fig:schematics}~(d)].
Upon excitation with the next pulse, a lesser number of monoclinic domains $N_\mathrm{mc}(F_1)$ is available for laser-induced SR transition.
At the higher excitation fluence, $F_2>F_1$, decrease of $N_\mathrm{mc}(F_2)$ in each heating-cooling cycle is even more pronounced.
Therefore, at each $T_0$, $\tilde N_\mathrm{mc}$ averaged over many excitation events appears to be lower at higher pump fluences.
This may account for the observed faster decrease of $\Theta_{PP(IP)}\sim \tilde N_\mathrm{mc}(F)$ with $T_0$ at higher fluences [Fig.~\ref{Fig:schematics}~(d)].
In contrast, in the case of excitation via change of $K_1$ in the cubic phase, heating-cooling cycles do not affect the initial state of the system [Fig.~\ref{Fig:schematics}~(f)]. 
As a result, increase of the pump fluence would result just in the increase of the detected amplitude $\Theta_{PP(IP)}\sim\Delta K_1(\Delta T)$ and would not affect the rate with which $\Theta_{PP(IP)}$ changes with $T_0$.}

We note, that the FMR linewidth in magnetite has been reported to narrow down as the equilibrium transition occurs concomitantly with decrease of the FMR frequency \cite{Srivastava-IEEE2020}.
However, in our pump-probe experiments only the damping time changes with fluence or temperature, indicating a different origin of the damping at low fluences and temperatures related to the spatial inhomogeneity of laser-induced SR transition under such conditions.
We also note that the laser-driven precession associated with the first-order transition, Morin point, has been reported in dysprosium orthoferrite and also showed nontrivial dependence of the precession damping on the laser pulse fluence \cite{Afanasiev_PRL2016}. 
However, in orthoferrites the damping of the precession was found to be correlated with the timescale of the transition in the orthoferrite, which is intrinsically limited to picosecond range and thus can be traced by precessional reorientation of the magnetization towards the new equilibrium axis.
In magnetite, timescale of the SR transition appears to be well below the resolution of our experiment and clearly happens faster than the half a period of the precession, $\sim20$~ps.
Given the subpicosecond timescale of the crystal and electronic structure switching \cite{DeJong-nat2013}, the initial change of the magnetic anisotropy can have similar speed.

\section{Conclusion}
In conclusions, we have demonstrated laser-induced magnetization precession in bulk single crystal magnetite, and have shown that the underlying mechanism of the excitation is the switching of anisotropy axis due to ultrafast SR transition. 
This expands our knowledge about response of this iron oxide to ultrafast optical excitation, along with earlier demonstrated insulator-metal and structural, i.e. Verwey, transition.
We found that SR transition can be triggered even by laser pulse with a fluence $F<1$~mJ/cm$^2$, i.e. below the threshold for the laser-induced heating required for material to overcome the transition temperature.
Furthermore, signatures of photo-induced SR transition are present in a wide range of initial sample temperatures both below and above transition temperature.
Joint detection of the precession via transient linear and quadratic  magneto-optical effects reveal that under such conditions the photo-induced transition occurs in separate domains, signifying the first-order character of it taking place also beyond the thermal equilibrium.
Revealing such broad range of conditions supporting photo-induced transitions was 
elusive in the transient reflectivity measurements.

Importantly, we show that there is a striking similarity in the threshold and saturation fluences and threshold sample temperatures for the Verwey and SR transition, suggesting their intrinsic link to the same driving mechanism. 
The latter for the case of Verwey transition was shown in to be a subpicosecond melting of specific charge and orbital ordering - trimerons, i.e. Fe$^{3+}$-Fe$^{2+}$-- Fe$^{3+}$ complexes \cite{DeJong-nat2013}.
Thus, our results sets up a basis for further studies aiming at revealing timescale of magnetic anisotropy change in Fe$_3$O$_4$.
Such research could rely, e.g., on joint X-ray study of femtosecond dynamics of spin and orbital momenta \cite{Boeglin-Nature2010}.
In our experiments, we used the ferromagnetic mode of the magnetization precession as a fingerprint of the photo-induced magnetic transition, which is intrinsically slow in low applied magnetic fields and thus hinders realistic timescale of the anisotropy change. 
Examining ultrafast SR in thin and ultrathin magnetite films readily available by a pulsed laser deposition technique \cite{Suturin-PRMater2018} may provide access to shorter timescales via observing higher-order laser-induced standing spin-wave modes \cite{Scherbakov-PRAppl2019}.

\begin{acknowledgments}
Single crystals of magnetite were grown by A.~M. Balbashov.  
We thank I.~V. Karpovskii for fruitful discussions. 
This work was done with partial financial support of the Russian Foundation for Basic Research (grant No.~20-02-00938).
\end{acknowledgments}

\appendix

\section{Calculation of the laser-induced heating}\label{app:heating_calculation}

\textcolor{revised}{Using data from the literature on latent heat $J_{\mathrm{L}}$~=~0.85~mJ/cm$^3$ \cite{latent_heat} and specific heat capacities for the metallic and insulating phases of magnetite \cite{heat_capacities}, the temperature increase $\Delta T$ and the resulting temperature $T_\mathrm{h}=T_0+\Delta T$ were calculated \cite{Mogunov_NatureComm2020} as a function of pump fluence $F$ (at $T_0=80$~K) and of the initial temperature of magnetite $T_{0}$ (at $F$=1.6 mJ/cm$^2$).}

\textcolor{revised}{For calculations, volumetric absorbed pump energy density $J$ was founded from the incident fluence as 
\begin{equation}
J=(1-R_\mathrm{i}) \alpha_i F,
\end{equation}
where $i=\left\{\mathrm{ins},\mathrm{met}\right\}$ stands for the insulating or metallic properties of the material at a particular $T_0$.
$R_\mathrm{ins}$ = 0.15 and $R_\mathrm{met}$ = 0.27 are a reflection coefficients and $\alpha_\mathrm{ins}\approx\alpha_\mathrm{met}$ = 0.72$\cdot$10$^{5}$ cm$^{-1}$ are the absorption coefficients of magnetite at the pump photon energy \cite{schlegel1979optical}.
Threshold $J_\mathrm{th}=1.3\cdot10^7$~J/m$^3$ and saturation energy densities $J_\mathrm{S}=2.7\cdot10^7$~J/m$^3$ are found from the corresponding fluences $F_\mathrm{th(S)}$.}

\textcolor{revised}{For the temperatures $T_0$ below $T_V$ and fluences below the threshold, $F<F_\mathrm{th}$, the temperature increase was calculated as
\begin{equation} 
T_\mathrm{h}=\frac{J}{c_\mathrm{ins}}+T_0,
\end{equation}
where 
$c_\mathrm{ins}=~0.8\cdot10^6$~J/(m$^3$K) is a volumetric heat capacity at $T_0=80$~K. 
For the pump fluences above the saturation threshold, $F>F_\mathrm{S}$, we consider that the absorbed pump energy firstly drives the phase transition ($J_\mathrm{th}+J_\mathrm{L}$) and the rest of it heats the material in the metallic phase:
\begin{equation} 
T_\mathrm{h}=\frac{(J-J_\mathrm{th}-J_\mathrm{L})}{c_\mathrm{met}}+\frac{J_\mathrm{th}}{c_\mathrm{ins}}+T_0,
\end{equation}
}
\textcolor{revised}{where $c_\mathrm{met}=~1.6\cdot10^6$~J/(m$^3$K) is a volumetric heat capacity at $T=123$~K.}

\textcolor{revised}{At the intermediate fluences $F_\mathrm{th}<F<F_\mathrm{S}$ both fractions of material, the one remaining in insulating phase and that experiencing the transition, contribute to heating. The temperature increase is found as \cite{Mogunov_NatureComm2020}:
\begin{equation} 
 T_\mathrm{h}=\Phi_\mathrm{ins} \frac{J_\mathrm{th}}{c_\mathrm{ins}} +\Phi_\mathrm{met} \frac{J_\mathrm{S}-J_\mathrm{th}-J_\mathrm{L}}{c_\mathrm{met}} +T_0,
\end{equation}
where}
\textcolor{revised}{\begin{equation} 
\Phi_\mathrm{ins}= \frac{1}{2}\left[1-\mathrm{erf}\left(\frac{J-J_\mathrm{th}}{\sigma_\mathrm{th}}\right) \mathrm{erf} \left( \frac{J-J_0}{\sigma} \right)\right],
\end{equation}
}
\textcolor{revised}{\begin{equation} 
\Phi_\mathrm{met}= \frac{1}{2}\left[1-\mathrm{erf}\left(\frac{J-J_\mathrm{S}}{\sigma_\mathrm{S}}\right) \mathrm{erf} \left( \frac{J-J_0}{\sigma} \right)\right],
\end{equation}
are weight factors of contributions from the insulator and metal phases.
In the calculations, values $\sigma$=$\sqrt{2}\cdot (J_\mathrm{S}-
J_\mathrm{th})/4$, $\sigma_\mathrm{S}$=$\sigma_\mathrm{met}$=$\sigma$/10, $J_0=(J_\mathrm{th}+J_\mathrm{S})/2$ were used.}

\begin{figure}
\includegraphics [width=\columnwidth] {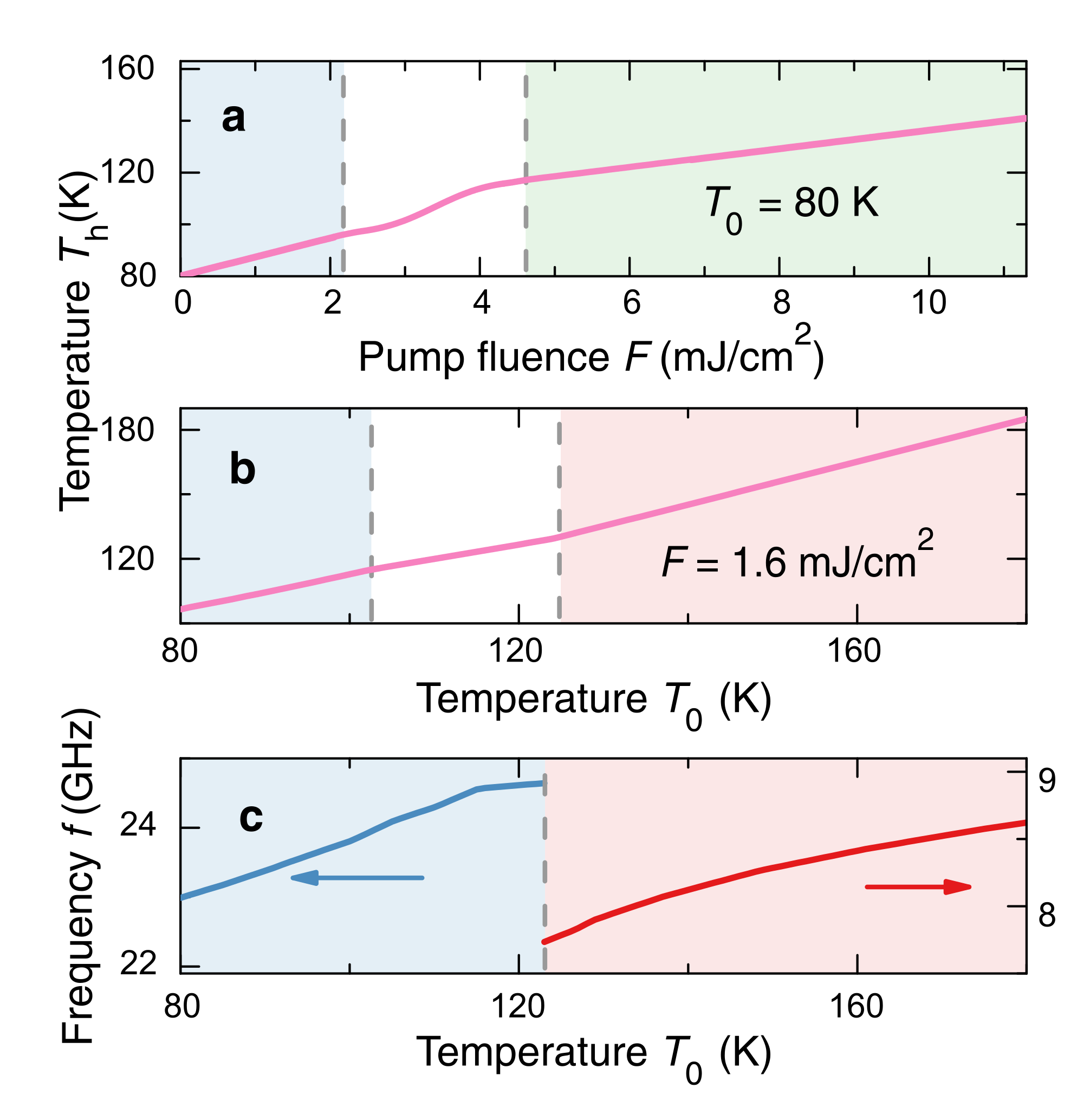}
\caption{\label{fig:Apendix} 
(a) Laser-induced heating temperature of the sample $T_\mathrm{h}$ as a function of  the pump fluence at fixed initial temperature $T_{0}=$~80~K.
(b) Laser-induced heating temperature of the sample $T_\mathrm{h}$ as a function of the initial temperature $T_{0}$ at fixed pump fluence $F=$~1.6~mJ/cm$^2$. 
(c) Calculated equilibrium FMR frequency as a function of the sample temperature.
}

\end{figure}

\textcolor{revised}{Figure \ref{fig:Apendix} (a) shows the dependence of the heating temperature $T_\mathrm{h}$ on the pump fluence at initial temperature $T_{0}=$~80~K  calculated using the formulas described above.}

\textcolor{revised}{Using dependences of the reflection coefficient $R(T_0)$ \cite{schlegel1979optical} and volumetric heat capacity on temperature, the heating of the magnetite was calculated with the same formulas for different temperatures at fixed pump fluence $F=$~1.6~mJ/cm$^2$ [Fig.~\ref{fig:Apendix}(b)]. 
In calculations we assumed that $F_\mathrm{th}$ linearly decreases from 2.2~mJ/cm$^2$ at $T_0=$80~K down to 1.6~mJ/cm$^2$ at $T_\mathrm{th}$.}

\section{Calculation of the FMR frequency} \label{app:FMR_calculation}

Ferromagnetic resonance frequency in magnetite at different temperatures was calculated with the Smit-Suhl approach \cite{gurevich1996magnetization}:

\begin{equation} \label{eq:Freq_Smit_Suhl}
  \omega=\frac{\gamma}{M_{0} \sin\theta_{0}}\sqrt{ U_{\theta \theta} U_{\varphi \varphi} - U^2_{\theta \varphi}}
\end{equation}
where the gyromagnetic ratio $\gamma$ is taken according to the reported value of the $g-$factor $g=2$ \cite{bickfordPRB1950}. $U_{i,j}=\partial^2U/\partial i \partial j$, $\{i,j\}= \{ \theta, \theta \} $, $ \{ \varphi, \varphi \}$, $\{ \theta, \varphi \} $ at equilibrium direction of $\mathbf{M} (\theta = \theta_0$ and $\varphi = \varphi_0 )$.

Below Verwey transition, $T<T_\mathrm{V}$, magnetite is a monoclinic crystal and has a uniaxial anisotropy.
At $T>T_\mathrm{V}$, magnetite structure is cubic and it possesses the corresponding magnetic anisotropy.
Magnetization dependent part of free energy is written as:
\begin{gather}
\label{eq:anisotropy_energy_trigon}
  \begin{aligned}
  U_\mathrm{uni}&= 0.5K_a  \sin^2{\theta}(\cos\varphi-\sin\theta)^2\\
  & + 0.5K_b  \sin^2\theta(\cos\varphi+\sin\theta)^2 + U_\mathrm{H},\quad\:
  T<T_\mathrm{V}\\
 U_\mathrm{c}&= K_1 \mathrm{sin}^2\theta(\mathrm{sin}^2\theta\;\mathrm{cos}^2\varphi\;\mathrm{sin}^2\varphi+\mathrm{cos}^2\theta)\\
  & + K_2\sin^4\theta\cos^2\theta\sin^2\varphi\cos^2\varphi + U_\mathrm{H},\quad
 T>T_\mathrm{V}\\
  \end{aligned}
\end{gather}

where $K_{a,b}$ are the leading uniaxial anisotropy parameters, $K_{1, 2}$ are the cubic anisotropy parameters, $U_\mathrm{H}$ is the Zeeman energy.
$\varphi$ and $\theta$ are azimuthal and polar angles of magnetization. 

Figure~\ref{fig:Apendix}~(c) shows the FMR frequencies as a function of the sample temperature calculated for the field $H=250$~mT applied along the $c([001])-$axis. 
As expected, the FMR frequency is at least twice as high in the low-temperature phase as compared to the high-temperature phase, in agreement with experimental data [Fig.~\ref{fig:Fig1}~(c,~d)].
We note some discrepancy between measured and calculated data, which may stem from somewhat different anisotropy parameters of the studied sample and a possible effect of the demagnetizing fields on the FMR frequency.
Nevertheless, calculations confirm the experimentally observed disappearance of the low-frequency FMR peak as the temperature becomes lower than $T_\mathrm{SR}$.

\bibliography{apssamp}

\end{document}